\documentclass[11pt,a4paper,reqno, aps, superscriptaddress, 
amssymb,amsmath,amsfonts]{amsart}
 
\usepackage[T1]{fontenc}
\usepackage{graphicx}
\usepackage[foot]{amsaddr}
\usepackage{caption}
\usepackage{dcolumn}
\usepackage{bm}
\usepackage{amssymb}
\usepackage[textheight=20 cm, textwidth=15cm, headheight=50pt, 
headsep=10pt, footskip=32pt]{geometry}
\usepackage{dsfont}
\usepackage{appendix}
\usepackage{subfig}
\usepackage{physics}
\usepackage{tikz}
\usepackage{amsmath}
\usepackage{amsthm}
\usepackage{orcidlink}
\usepackage{subcaption}
\usepackage{mathtools}
\usepackage[normalem]{ulem} 
\usepackage[colorinlistoftodos]{todonotes}

\newcommand{\Id}{{\rm 1\hspace{-0.9mm}l}}

\usepackage{diagbox}
\usepackage{booktabs}

\newcommand{\proj}[1]{\ensuremath{\ketbra{#1}{#1}}}

\title[Multiple-shot unitary channels discrimination]{Experimental study of 
multiple-shot unitary channels discrimination using the IBM Q computers}

\author{Adam Bílek$^{1,2}$\orcidlink{0000-0001-8380-5338}}
\author{Jan Hlisnikovský$^{*,1,2}$\orcidlink{0009-0001-3554-7167}}
\author{Tomáš Bezděk$^1$\orcidlink{0009-0006-3021-1855}}
\author{Ryszard Kukulski$^2$\orcidlink{0000-0002-9171-1734}}
\author{Paulina Lewandowska$^2$\orcidlink{0000-0003-1564-7782}}
\address{$^1$ Department of Applied Mathematics, Faculty of Electrical 
Engineering and Computer science, VSB-Technical University of Ostrava, 
17.~listopadu 2172/15, 708 33 Ostrava, Czech Republic}
\address{$^2$IT4Innovations, VSB~-~Technical University of Ostrava, 
17.~listopadu 2172/15, 708 33 Ostrava, Czech Republic}
\address{$^*$ \textit{Corresponding author. E-mail: 
\texttt{jan.hlisnikovsky@live.com}}} 

\begin{document}
\maketitle
\section{Abstract}
Tasks involving black boxes appear frequently in quantum computer science. 
An example that has been deeply studied is quantum channel discrimination. 
In this work, we study the discrimination between two quantum unitary 
channels in the multiple-shot scenario. We challenge the theoretical 
results 
concerning the probability of correct discrimination with the results 
collected from experiments performed on the IBM Quantum processor Brisbane. 
Our analysis shows that neither too deep quantum circuits nor circuits that 
create too much entanglement are suitable for the discrimination task. We 
conclude that circuit architectures which minimize entanglement overhead 
while preserving discrimination power are significantly more resilient to 
hardware noise if their depth does not overpass threshold value.

\section{Introduction}
In the last decade, quantum computing has become a reality. Quantum 
algorithms of increasing degree of complexity are currently being 
implemented on more and more complex quantum devices. In effect, 
high-quality solutions to some real-world problems are expected to be 
forthcoming soon. This situation motivates the need for the certification 
and benchmarking of various quantum devices \cite{eisert2020quantum, 
mckay2023benchmarking, proctor2025benchmarking}. The discrimination task of 
quantum operators constitutes one of the certification methods and quality 
metrics for benchmarking quantum architectures 
\cite{jalowiecki2023pyqbench}. The theoretical background of the quantum 
discrimination task has been widely developed.  The primary task of 
discrimination involves a one-shot scenario of discrimination between 
quantum operators. We can imagine that we have an unknown quantum device, a 
black-box. The only information we have is that it performs one of two 
quantum operators, say $\mathcal{T}$ and $\mathcal{S}$. Our goal is 
two-fold. First, we want to determine the highest possible probability of 
correct guessing. Secondly, we need to devise an optimal strategy that 
maximizes the probability of success.

 The problem of single-shot discrimination of quantum states was solved 
analytically by Helstrom a few decades ago in~\cite{helstrom1969quantum, 
watrous2018theory}. The authors have calculated the probability of correct 
discrimination between two quantum states using the notion of a trace norm. 
Next, there are many modifications of the origin problem for quantum states 
that were also considered \cite{sen2024incompatibility, 
bandyopadhyay2018optimal, calsamiglia2010local, bergou200411, 
qiu2008minimum}. For the discrimination tasks of the quantum channels 
\cite{watrous2018theory}, the probability of success of the discrimination 
can be formulated by the diamond norm \cite{watrous2013simpler, 
benenti2010computing}, which can be computed by semidefinite programming 
\cite{ramana1997exact, watrous2013simpler}. However, for increasing the 
dimension of quantum channels in the diamond norm, the size of the problem 
does not allow efficient computation \cite{Gawron2018}. In general, for 
general quantum channels, entanglement is necessary for optimal 
discrimination \cite{watrous2018theory}. The exception is, for example, the 
discrimination task between unitary channels\cite{watrous2018theory, 
ziman2010single}. Furthermore, the probability of correct discrimination 
between two unitary channels can be expressed in the notion of the 
numerical 
range \cite{nr}. The discrimination tasks for general quantum measurements, 
von Neumann measurements, or SIC POVMs were also considered in the 
literature \cite{sedlak2014optimal, krawiec2024discrimination, 
puchala2018strategies, krawiec2020discrimination, ji2006identification, 
cao2015perfect}. Lastly, the theory of an indefinite causal structure is 
one 
of the attractive topics. 
 This approach uses the notion of process matrices 
\cite{oreshkov2012quantum, araujo2015witnessing}, which can be seen as a 
generalization of quantum combs. The single-shot discrimination of the 
process matrices was introduced \cite{lewandowska2023strategies}. 
 
What if we have multiple copies of quantum operators? 
The most general discrimination approach is known as adaptive strategy 
\cite{salek2022usefulness}. This strategy can be described by using the 
term 
quantum combs. Here, we assume freedom in choosing the setting of quantum 
operations and any processing between them. In addition, numerical 
investigation of quantum channel discrimination showed that using 
indefinite 
causal order, we can achieve greater success discrimination 
\cite{bavaresco2021strict, bavaresco2022unitary}. However, determining an 
optimal strategy to realize such tasks is a significant challenge in 
practice. This is why we often limit ourselves to parallel 
\cite{duan2016parallel} and sequential \cite{chiribella2008memory} schemes. 
The parallel scheme assumes N copies of a given quantum operation 
distributed over N entangled systems and final measurement with no 
additional processing, whereas
the sequential scheme assumes the implementation of N copies of the quantum 
operation on one system, allowing additional processing between them. 

 One of the first theoretical results was the study of discrimination of 
multipartite unitary operations \cite{bavaresco2021strict}. The authors 
were 
shown that perfect discrimination can be achieved using a parallel scheme 
\cite{duan2016parallel}. Simultaneously, it is also possible to create a 
perfect discrimination scheme of unitary channels without creating the 
entanglement with the auxiliary system \cite{duan2007entanglement}. It can 
be done by using the sequential approach and specific processing at the end 
of the circuit. On the other hand, for general quantum channels, neither 
parallel nor sequential strategy guarantees optimal solutions 
\cite{harrow2010adaptive, hayashi2009discrimination}.  The advantage of 
adaptive strategies became apparent for a task concerning 
entanglement-breaking channels \cite{harrow2010adaptive}.  The 
discrimination tasks in which adaptive scenarios were also investigated in 
\cite{chiribella2008memory}, \cite{wang2006unambiguous, 
krawiec2020discrimination}. In the work \cite{duan2009perfect}, the authors 
have formulated the necessary and sufficient conditions under which quantum 
channels can be perfectly discriminated, while in 
\cite{krawiec2020discrimination}, the authors have formulated conditions 
for 
the perfect discrimination of two measurements. Furthermore, in 
\cite{puchala2021multiple}, the authors have shown that any possible 
adaptive method does not offer any advantage over the parallel scheme for 
von Neumann measurements. 
  
One of the first experimental results for the parallel and sequential 
discrimination scheme of unitary channels was presented 
\cite{liu2019distinguishing}.
In the era of NISQ devices, we need to take into account certain 
limitations.  Due to the existing decoherence, sequential schemes could not 
be practical for larger numbers of copies. At the same time, parallel 
schemes for NISQ architectures are also not possible to implement because 
of 
the limited number of qubits. 
 Perhaps intermediate schemes called sequentially-paralleled will prove to 
solve both of these obstacles.
 This scheme assumes that we have $N = w\cdot d$, copies of a quantum 
operator, where $k$ and $l$ are natural numbers.  The 
sequentially-paralleled scheme of width $w$ and depth $d$ consists of $d$ 
applications of $w$ copies of the quantum operator applied simultaneously 
to 
the quantum state. The initial state $\rho$ evolves through these layers, 
combining parallelism within each layer and a sequential structure across 
layers. In general, one could consider incorporating additional 
intermediate 
processing operations between layers. 

In this work, we present a comprehensive study of various scenarios of 
multiple-shot discrimination of quantum unitary channels. We consider 
parallel, sequential, and sequentially-paralleled quantum networks for 
discrimination between two qubit unitary channels. We challenge the 
theoretical results concerning the probability of correct discrimination 
with the results collected from experiments performed on the IBM Quantum 
processor Brisbane. Based on several examples, the analysis shows that 
neither too deep quantum circuits nor circuits that create too much 
entanglement are suitable for the discrimination task.

This paper is organized as follows. Section \ref{sec:preli} presents the 
necessary mathematical preliminaries. Section \ref{sec:disc} introduces the 
task of discriminating quantum channels in single and multiple-shot 
scenarios. 
Sections \ref{sec:reali-1} and \ref{sec:reali2} are dedicated examples of 
quantum unitary channel discrimination performed on NISQ devices. Finally, 
Section \ref{sec:conclusion} summarizes the main contributions and results 
of this work.

\section{Mathematical preliminaries}
\label{sec:preli}

Let $\mathcal{X}$ be a complex Euclidean space, then L($\mathcal{X}$) 
denotes the collection of all linear mappings of the form $A: \mathcal{X} 
\rightarrow \mathcal{X}$. An operator $X\in \text{L}(\mathcal{X})$ is 
positive semi-definite if $\bra{x}X\ket{x} \geq 0$ for all $\ket{x} \in 
\mathcal{X}$. The set of all such operators is written as $\text{Pos} 
(\mathcal{X})$.  By \( \Omega(\mathcal{X}) \) we denote the set of quantum 
states $\rho$ $\in \text{Pos}(\mathcal{X})$ such that $\text{Tr}\rho=1$. 
Let 
$X\in \text{L}(\mathcal{X})$, by $\text{spec(X)}$ we express the set of all 
eigenvalues of $X$.
 
 The set of all linear mappings from L($\mathcal{X}$) into 
L($\mathcal{Y}$), $\Phi:\text{L}(\mathcal{X})  \rightarrow 
\text{L}(\mathcal{Y})$, will be denoted as T($\mathcal{X}$,$\mathcal{Y}$). 
 A linear map $\Phi \in \text{T}(\mathcal{X}$,$\mathcal{Y})$  is positive 
if holds $\Phi(P) \in \text{Pos}(\mathcal{Y})$ for all $P \in 
\text{Pos}(\mathcal{X})$, whereas $\Phi$ is a completely positive map (CP) 
if $\Phi \otimes \Id_{\text{L}(\mathcal{Z})}$ is a positive map for every 
complex Euclidean space $\mathcal{Z}$. We say $\Phi$ is trace-preserving 
(TP) if it holds that Tr($\Phi(X)$) = Tr($X$) for all $X 
\in\text{L}(\mathcal{X})$. A linear map $\Phi \in 
\text{T}(\mathcal{X},\mathcal{Y})$, which is a CPTP map, is called a 
quantum 
channel. The collection of all quantum channels is denoted as 
$\mathrm{C}(\mathcal{X},\mathcal{Y})$ (with a shortcut 
$\mathrm{C}(\mathcal{X}) \coloneqq \mathrm{C}(\mathcal{X},\mathcal{X})$). 
Next, we distinguish a special subset of quantum channels known as unitary 
channels, $\Phi_U \in \mathrm{C}(\mathcal{X})$ defined as $\Phi_U(X) = 
UXU^{\dagger}$ where $U \in\text{L}(\mathcal{X})$ is the unitary matrix.
 
 A Positive Operator-Valued Measure (POVM) $\mathcal{P}$ is a collection of 
operators, the so-called effects $\{E_0, \cdots, E_n\} \subset 
\text{Pos}(\mathcal{X})$ with the property of $\sum_{i=0}^{n} E_i = \Id$. 
According to the Born rule for a given quantum state $\rho$ the probability 
of obtaining the result $E_i$ is given by $\text{Tr}(E_i\rho)$.

 An useful tool for studying the discrimination of unitary channels is the 
concept of the numerical range of an operator. For $X \in 
\text{L}(\mathcal{X})$ we define the numerical range of ${X}$ as the set 
\begin{equation}
    \text{W}(X) \coloneqq \{ \bra{x}X\ket{x}:\ket{x}\in\mathcal{X}, \langle 
x | x \rangle = 1\}.
\end{equation}
The Hausdorff-Töplitz Theorem \cite{Hausdorff1919,Toeplitz1918} states that 
\text{W}($X$) is a convex set. If $X$ is normal, the numerical range is a 
convex hull of its eigenvalues. For unitary matrices $U$ we define the arc 
function $\theta(U)$ as the length of the smallest arc on the unit circle 
that contains all the eigenvalues of the unitary operator $U$. 
Mathematically, this can be expressed as
\begin{equation}
\begin{split}
     \theta(U) \coloneqq & \min \Bigl\{ \Delta \in [0,2\pi) \, :\, 
\exists\, \alpha \in [0,2\pi) \text{ such that }  \\ & \text{spec}(U)  
\subset \{e^{i\theta} : \theta \in [\alpha, \alpha+\Delta]\} \Bigr\}.
    \end{split}
\end{equation}
Lastly, let us introduce the diamond norm in the space 
$\text{T}(\mathcal{X},\mathcal{Y})$. For $\Phi\in 
\text{T}(\mathcal{X},\mathcal{Y})$ it is defined as
\begin{equation}
\| \Phi \|_{\diamond} = \left\| \Phi \otimes \Id_{\text{L}(X)} 
\right\|_1,    
\end{equation}
where $\| \Phi \|_1 = \max \{ \| \Phi(X) \|_1 : X \in 
\text{L}(\mathcal{X}), \| X \|_1 \leq 1 \}$ and $ \| Y\|_1$ is Schatten 
1-norm of $Y \in \text{L}(\mathcal{Y})$.

\section{Discrimination of quantum channels}\label{sec:disc}

We will consider the following scenario of quantum channel discrimination. 
Suppose that we have a classical description of two quantum channels, 
$\Phi_0, \Phi_1 \in \mathrm{C}(\mathcal{X}, \mathcal{Y})$ and a black box 
containing a quantum channel $\Phi$, which is either $\Phi_0$ or $\Phi_1$. 
We would like to determine whether $\Phi = \Phi_0$ or $\Phi = \Phi_1$ have 
been hidden in the black box. To reveal the value of $\Phi$ we can 
construct 
a quantum experiment which will consist of an initial state $\rho \in 
\Omega(\mathcal{X} \otimes \mathcal{Z})$ and a binary measurement $\{E_0, 
E_1\} \subset \text{Pos}(\mathcal{Y} \otimes \mathcal{Z})$. The channel 
$\Phi \otimes \Id_{\text{L}(\mathcal{Z})}$ is applied on $\rho$ and then 
the 
output state $(\Phi \otimes \Id_{\text{L}(\mathcal{Z})})(\rho)$ is measured 
by $\{E_0, E_1\}$. The measurement label defines our guess about the hidden 
value of $\Phi$ (the label $0$ associated with the effect $E_0$ indicates 
$\Phi = \Phi_0$). In such setup, the probability $p_{\text{succ}}$ of 
successful channel discrimination is given as
\begin{equation}
    p_{\text{succ}} = \frac12 \text{Tr}(E_0(\Phi_0 \otimes 
\Id_{\text{L}(\mathcal{Z})})(\rho)) + \frac12 \text{Tr}(E_1(\Phi_1 \otimes 
\Id_{\text{L}(\mathcal{Z})})(\rho)).
\end{equation}

The goal of our task is to construct $\rho$ and $\{E_0, E_1\}$ that 
maximize $p_{\text{succ}}$. The auxiliary system $\mathcal{Z}$ is of 
arbitrary size and provides a resource to increase the probability of 
success. From the Holevo-Helstrom theorem \cite{holevo1973statistical, 
helstrom1969quantum} we know the probability of successful channel 
discrimination can be expressed by the help of the diamond norm
\begin{equation}
    p_{\text{succ}} = \frac{1}{2} + \frac{1}{4} \left\|  \Phi_0 - \Phi_1 
\right\|_\diamond.
\end{equation}

\subsection{Single-shot discrimination of quantum unitary channels}

 Considering the task of discrimination between unitary channels  $ 
\Phi_U,\Phi_V  $,  the diamond norm between such channels can be expressed 
using the notion of the numerical range as \cite{watrous, Watrous2011}
\begin{equation}
    ||\Phi_{U}-\Phi_{V}||_{\diamond}=2\sqrt{1-\nu^{2}},
\end{equation}
where $\nu = \min_{w \in W(V^\dagger U)} |w|$.
From the above proposition it follows that unitary
channels $\Phi_U$ and $\Phi_V$ are perfectly distinguishable if and only
if $0 \in W(V^\dagger U)$. The above can also be formulated as there exists 
a density matrix $\sigma$, such that $\tr (V^\dagger U \sigma) = 0$.

Using the results of \cite{acin2001statistical, d2001using, 
duan2007entanglement} we can calculate $\nu$ using the arc function 
$\theta(V^\dagger U)$, determining the condition for perfect discrimination 
between $U$ and $V$ in single-shot scenario. We achieve perfect 
discrimination if and only if
\begin{equation}
    \theta(V^\dagger U) \geq \pi.
\end{equation}
This condition ensures that zero lies in the numerical range of $V^\dagger 
U$. Hence, an input state can be chosen such that it evolves under $\Phi_U$ 
to a state that is orthogonal to the same input state evolved under 
$\Phi_V$.

\subsection{Multiple-shot discrimination of unitary channels}

When multiple uses of the unitary channel are available, perfect 
discrimination may be achieved even if the single-shot condition 
$\theta(V^\dagger U) \geq \pi$ is not met \cite{acin2001statistical}. The 
theoretically best strategy in that case is a parallel scheme which 
consists 
of $N$ copies of the unitary $U$ applied simultaneously to $N$ quantum 
registers. In that case, the problem is to discriminate identity 
$U^{\otimes 
N}$ and unitary matrix $V^{\otimes N}$. Since the arc function satisfies 
the 
scaling relation $\theta((V^{\otimes N})^\dagger U^{\otimes N}) = N 
\theta(V^\dagger U)$, for $N\theta(V^\dagger U)<2 \pi$, then the condition 
for perfect discrimination is $N \theta(V^\dagger U) \geq \pi$ 
\cite{duan2007entanglement}. This leads to the condition on the minimum 
number of copies required for perfect discrimination is given by
\begin{equation}
    N \geq \left\lceil \frac{\pi}{\theta(V^\dagger U)} \right\rceil.
\end{equation}
Thus, even when $\theta(V^\dagger U) < \pi$, as long as $\theta(V^\dagger 
U) > 0$, perfect discrimination can still be achieved by using a sufficient 
number of copies of $U$ and $V$.

\subsection{Experimental set-up}
A wide range of multi-shot discrimination strategies have been explored in 
the literature. In this work, we restrict our attention to three 
representative classes: parallel schemes, sequential schemes, and 
rectangular hybrid schemes that interpolate between these two. Experimental 
comparison of the performance of these strategies will constitute the 
contribution of our work. Our attention is put on the discrimination of 
qubit unitary channels $U$ and $V$. For simplicity we will consider only 
the 
cases that satisfy
\begin{equation}
    \theta(V^\dagger U) =  \frac{\pi}{N}.
\end{equation}

Let $N = wd$ for $w,d \in \mathbb{N}$. In our set-up $N$ copies of the 
black box $\Phi$ can be composed in the rectangular shape spread on $w$ 
qubits with the depth of the circuit $d$. On each qubit the unknown 
operation is composed $d$ times, namely we have
\begin{equation}
     \Phi^{\otimes w} \circ\Phi_{X_{d-1}} \circ \Phi^{\otimes w} \circ 
\cdots\circ \Phi_{X_1} \circ \Phi^{\otimes w},
\end{equation}
where $X_1,\ldots,X_{d-1}$ are arbitrary unitary matrices defined on $w$ 
qubits which are responsible for mid-circuit processing. In special cases, 
when $w=N$ and $d=1$ we have the parallel scheme $\Phi^{\otimes N}$ (no 
processing is needed) and when $w=1$ and $d=N$ we have the sequential 
scheme 
$\Phi\Phi_{X_{d-1}}\cdots\Phi_{X_1}\Phi$.

In general, for $d>1$ if we do not apply mid-circuit processing ($X_i=\Id$) 
then theoretically the scheme will be strictly worse than the parallel 
scheme ($d=1$). For example, comparing sequential scheme without processing 
($d=N$) and parallel scheme, we get $\theta((V^N)^\dagger U^N) < 
\theta((V^{\otimes N})^\dagger U^{\otimes N}) = \pi$ for most (with respect 
to the Haar measure) pairs of $U,V$. However, if we choose a simple 
processing $X_i = (V^\dagger)^{\otimes w}$ we get
\begin{equation}
    \theta(((V^\dagger U)^d)^{\otimes w} )=w\theta((V^\dagger U)^d 
)=wd\theta(V^\dagger U )=N \frac{\pi}{N}=\pi.
\end{equation}
This means that for all $w,d$ satisfying $wd = N$ we achieve perfect 
discrimination for all $N$-shot scenarios we consider. Hence, in theory, 
all 
three scenarios: parallel, sequential, and hybrid are equal and give 
$p_{\text{succ}} = 1$. 

The only remaining question is how to find optimal $\rho$ and $\{E_0, 
E_1\}$? The precise form of these variables depends on $U,V$ and will be 
calculated later for each pair of unitary channels considered. It is worth 
sympathizing at this point that there is no need to use an auxiliary system 
$\mathcal{Z}$ for the discrimination of quantum unitary 
channels~\cite{bavaresco2022unitary}. 

\section{Discrimination of unitary channels on IBM-Q with no 
processing}\label{sec:reali-1}

This section is divided into four main parts. 
At the beginning, we will discuss in details the components of 
discrimination schemes, and later on their decompositions into native 
gates. 
Next, we will talk about transpilation approaches, and finally we will 
present the results obtained from the IBM Q device. All experiments were 
executed on the IBM Quantum processor Brisbane.

\subsection{Example 1}
In this example, we will distinguish between identity $\Phi_{\Id}$ and 
$\Phi_{\text{RZ}(\phi)}$ for \text{RZ}$(\phi) = \begin{pmatrix} 
e^{-i\frac{\phi}{2}} & 0\\ 0 & e^{i\frac{\phi}{2}} \end{pmatrix}$ without 
processing between the particular application of the unitary channel ($X_i 
= 
\Id$ for each $i=1,\ldots,d$). 
Let $N = wd$ for $w,d \in \mathbb{N}$. In our setup $N$ copies of the black 
box $\Phi$ can be composed in the rectangular shape spread on $w$ qubits 
with the depth of the circuit $d$. On each qubit the unknown operation is 
composed $d$ times, so namely we have
\begin{equation}
    \underbrace{\Phi^d \otimes \cdots \otimes \Phi^d}_{w}.
\end{equation}
To determine a discriminator $\ket{\psi}$, we need to find a unit vector 
satisfying \begin{equation}
    \bra{\psi} \text{RZ}(d \phi )^{\otimes w} \ket{\psi} = 0.
\end{equation} The most distant pair of eigenvalues of $\text{RZ}(d \phi 
)^{\otimes w}$ are $-i, i$ corresponding to eigenvectors $\ket{0\cdots0}, 
\ket{1\cdots1}$, respectively. Hence, we can show that 
\begin{equation}
    \ket{\psi} = \frac{1}{\sqrt{2}}\left(\ket{0\cdots0} + \lambda 
\ket{1\cdots1} \right) \in \mathbb{C}^{2^w},
\end{equation}
for any unit number $\lambda \in \mathbb{C}$. If $\Phi = \Phi_\Id$, the 
output state is equal to $\ket{\psi_0} = \ket{\psi}$. Otherwise, if $\Phi = 
\Phi_{\text{RZ}(\phi)}$, then the output states are equal to $\ket{\psi_1} 
= 
\text{RZ}(d \phi )^{\otimes w}\ket{\psi} = 
\frac{-i}{\sqrt{2}}\left(\ket{0\cdots0} -\lambda \ket{1\cdots1} \right)$.
As $\ket{\psi_0}$ and $\ket{\psi_1}$ are orthogonal, we can find $E_0 \ge 
\proj{\psi_0}$ and $E_1 \ge \proj{\psi_1}$ (in particular $E_0 = 
\proj{\psi_0}$ and $E_1 = \Id - E_0$), which in summary will guarantee 
$p_{\text{succ}} = \frac12 \text{Tr}(E_0(\Phi_{\Id^{\otimes 
w}})(\proj{\psi})) + \frac12 \text{Tr}(E_1(\Phi_{\text{RZ}(d \phi 
)^{\otimes 
w}})(\proj{\psi})) = \frac12 \text{Tr}(E_0 \proj{\psi_0}) + \frac12 
\text{Tr}(E_1\proj{\psi_1}) = 1$.

\subsection{Components of the discrimination scheme and decompositions}
We divided the implementation of our discrimination circuit into two 
distinct components: the discriminator and the measurement. The unknown 
unitary gates, representing the quantum channel to be identified, are 
inserted between these two parts. In practical experiments, we focus 
specifically on schemes that in theory achieve perfect discrimination 
between the identity operation and the $\text{RZ}(\phi)$ gate. To ensure 
that the condition previously discussed is met, we set the angle $\phi$ to 
$\pi / N$, where $N$ denotes the total number of uses of the unknown gate 
in 
the circuit. This value is determined by the product of the width (the 
number of qubits used in parallel) and depth (the number of successive 
applications of the unknown gate).

\begin{figure}[!h]

\begin{minipage}{0.45\textwidth}
\centering
\includegraphics[height=0.25\textheight  ]{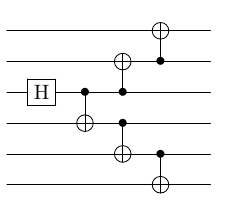}
\caption{CNOT implementation of the discriminator.}
\label{circ:cnot_disc}
\end{minipage}
\hfill
\begin{minipage}{0.45\textwidth}
\centering
\includegraphics[height=0.25\textheight  ]{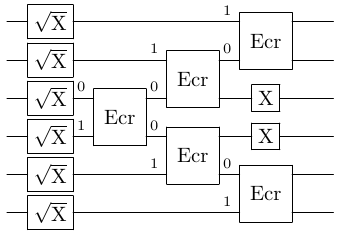}
\caption{ECR implementation of the discriminator.}
\label{circ:ecr_disc}
\end{minipage}
\end{figure}

After the usage of the discriminator, the circuit has to be in a maximal 
entangled state, on $N$ qubits in the form
\begin{equation}
    \ket{\psi} = \frac{1}{\sqrt{2}} \left( \ket{0}^{\otimes N} + \alpha 
\ket{1}^{\otimes N} \right).
\end{equation}
To achieve this, we used a cascade of CNOT gates, as can be seen for six 
qubits in Fig.~\ref{circ:cnot_disc}. 
This discriminator is created based on the obvious pattern that is commonly 
used to create the GHZ state \cite{eldredge2018optimal}.

 We run our experiments on IBM Brisbane, where the CNOT gate is not a 
native gate. To reduce the number of gates in the circuit, we prepare the 
discriminator specially designed for Eagle R3 architecture \cite{r3}. Then, 
we use ECR gates to create entanglement between qubits, omitting unrolling 
of CNOT gates during transpilation at the same time. 
The first part of the discriminator consists of SX gates on all qubits and 
then a cascade of ECR gates of similar structure as the CNOT cascade in the 
first case.
Unlike the discriminator based on CNOT gates, we also had to add several X 
gates to the end of the discriminator to get the desired quantum state.
There is little to no pattern in the qubits on which the X gate has to be 
applied. Therefore, we had to prepare the discriminator for each number of 
qubits separately by hand.
In Fig.~\ref{circ:ecr_disc}, we see the decomposition of ECR gate on six 
qubits. As we could see, in this case the X gates are applied on second and 
third qubits. 

\begin{figure}[!h]
\begin{minipage}{0.44\textwidth}
\centering
\includegraphics[height=0.3\textheight  
]{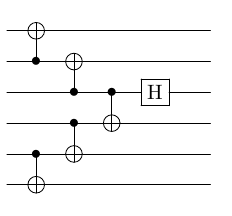}
\caption{CNOT implementation of the short measurement.}
\label{circ:cnot_simpl_measurement}
\end{minipage}
\hfill
\begin{minipage}{0.55\textwidth}
\centering
\includegraphics[height=0.3\textheight  ]{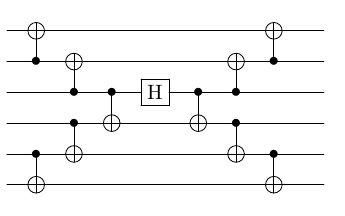}
\caption{CNOT implementation of the XOR measurement.}

\label{circ:cnot_xor_measurement}
\end{minipage}
\end{figure}

\begin{figure}[!h]
\centering
\includegraphics[height=0.3\textheight ]{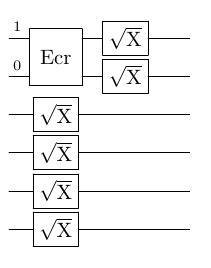}
\caption{ECR implementation of the short measurement.}
\label{circ:ecr_simpl_measurement}
\end{figure}

Two distinct methods were used to implement a circuit representing the 
measurement. The first method, referred to as the short measurement 
approach, is characterized by a reduced circuit depth and overall gate 
count. However, this advantage is offset by an increased susceptibility to 
bit-flip errors.
This circuit could be implemented using CNOT gates, presented in Fig. 
\ref{circ:cnot_simpl_measurement}, or ECR gates, as in Fig. 
\ref{circ:ecr_simpl_measurement}.
For the CNOT-based implementation, the measurement outcomes exhibit a 
regular structure: the scheme for the identity channel consistently yields 
the all-zero bitstring, while for the unitary channel 
$\Phi_{\text{RZ}(\theta)}$ produces bitstrings that are zero in all 
positions except for a single qubit set to one. The second measurement 
approach is different; it is much deeper, but the result for the identity 
channel consists of all zeros, and the result for the unitary channel 
consists of all ones.
This enables omitting several bit flip errors; we take a measured result, 
and if the majority of bits have value zero, we take it as the result of 
all 
zeros. In the case of a six-qubit system, and in the absence of noise, the 
observed bitstrings are $000000$ for identity and $001000$ for the unitary 
channel $\Phi_{\text{RZ}(\theta)}$.

In contrast, the ECR-based implementation returns a more complex 
distribution of output bitstrings, lacking the clear structure observed in 
the CNOT-based case.
With this implementation, we get two disjoint subsets of all possible 
results, one subset for identity and the second subset for the unitary 
channel $\Phi_{\text{RZ}(\theta)}$. 
For our example on six-qubit system, we obtain the following two sets of 
bitstrings after measurement: $\{001111, 010111, 101101, $ $110101, 100001, 
100111,$$ 011101, 110011,$ $000011, 111001,$$ 011011, 111111, $ $ 001001, 
000101, $ $010001, $ $101011\}$ for the identity channel and $\{110001, 
001011, 
011111,$$ 100011, 001101, 110111,$ $ 000001,$ $ 010101, 000111, 101001, 
111101, 010011,101111, 111011, 011001, $ $ 100101\}$ for 
$\Phi_{\text{RZ}(\frac{\pi}{6})}$.
The bitstring 001111 from the first set and 001101 from the second differ 
by only one bit, which may be problematic in a noisy environment.
Even in this small example, we could see that processing these results is 
not so straightforward but is compensated for by the simplicity of the 
measurement circuit.

In both measurement schemes, if it is not possible to determine which 
channel was applied during experiment, i.e., when the distance to both 
candidate channels is equal, then the output is assigned randomly via a 
coin 
flip. This approach ensures that each circuit run yields a definite 
outcome, 
thereby avoiding missing data points.
Analogically, if the majority of bits have value one, we take this as the 
result of all of them.

\begin{figure}
\centering
\includegraphics[height=0.3\textheight  ]{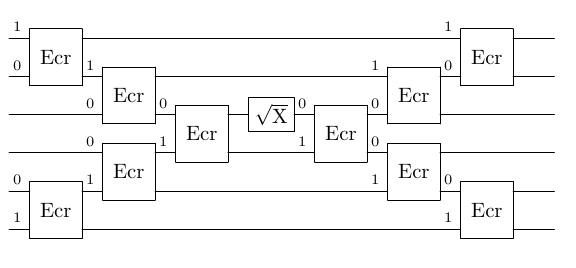}
\caption{ECR implementation of the XOR measurement.}
\label{circ:ecr_xor_measurement}
\end{figure}

The ECR gates are oriented, so we cannot swap qubits that are being acted 
on by ECR.
This fact causes that simple mapping logical to physical qubits on IBM 
Eagle R3 \cite{r3}, without transpilation of our ECR discriminator and 
measurement, is possible just for circuits not bigger than few qubits.
For discrimination tasks for higher dimentions, we have two possible 
solutions. For slightly bigger circuits, we are able to manually rearrange 
the circuit, but it has to be done specifically for every number of qubits. 
The second possibility is to use swap gates. This enables relatively easy 
mapping of the circuits, but it enlarges the depth of the circuits.

\subsection{Comparative analysis of transpilation}
To evaluate the efficacy of different quantum circuit transpilation 
approaches, particularly concerning the choice between CNOT and ECR gates 
and the impact of manual qubit mapping, experiments were conducted on the 
$6$-qubit and the $11$-qubit systems. The objective is to compare the 
performance, measured by result accuracy, for circuits utilizing either the 
short measurement or the XOR-based measurement scheme, while varying the 
transpilation method. 
In all these cases, a pure parallel discrimination scheme was employed.

\subsubsection{6-qubit system evaluation}

For the 6-qubit configuration, four distinct transpilation strategies were 
evaluated for the short measurement and XOR-based measurement protocols:
\begin{enumerate}
    \item \textbf{CNOT + Transpiler:} Circuit implemented using CNOT basis 
gates, processed by the Qiskit transpiler with optimization level 3.
    \item \textbf{ECR + Transpiler:} Circuit implemented using ECR basis 
gates, processed by the Qiskit transpiler with optimization level 3.
    \item \textbf{ECR + Transpiler + Fixed Mapping:} Circuit implemented 
using ECR basis gates, processed by the Qiskit transpiler with optimization 
level 3 and a predetermined, fixed mapping of logical qubits to physical 
qubits.
    \item \textbf{ECR + Fixed Mapping (No Opt.):} Circuit implemented using 
ECR gates, utilizing a fixed logical-to-physical qubit mapping, bypassing 
subsequent transpiler optimization passes.
\end{enumerate}

The experimental results, summarized in Table 
\ref{table:transpilation_6_qubits}, indicate that for this system size, 
applying a fixed qubit mapping does not yield a discernible improvement in 
accuracy compared to relying solely on the transpiler's default ECR 
implementation. This observation is potentially attributable to the limited 
circuit depth and complexity inherent in six-qubit systems, where the 
transpiler's optimization might already find near-optimal solutions without 
explicit mapping constraints.

\begin{table}[h!]
\centering
\begin{tabular}{|l| c | c |}
\hline
\diagbox{Transpilation Strategy}{Measurement} & Short & XOR \\ \hline
CNOT + Transpiler & $88.8\%$ & $86.4\%$ \\ \hline
ECR + Transpiler & $83.8\%$ & $90.0\%$ \\ \hline
ECR + Transpiler + Fixed Map & $84.4\%$ & $85.3\%$ \\ \hline
ECR + Fixed Map (No Opt.) & $83.3\%$ & $85.6\%$ \\ \hline
\end{tabular}
\caption{Accuracy of different transpilation strategies for discrimination 
scheme on $6$-qubit obtained on IBM Brisbane, using short and XOR 
measurement schemes. Each circuit was executed with 10,000 shots. Ambiguous 
measurement outcomes were randomly assigned.}
\label{table:transpilation_6_qubits}
\end{table}

\subsubsection{11-qubit system evaluation}

A subsequent set of experiments was performed using the $11$-qubit system. 
Five transpilation strategies were assessed for each measurement type:
\begin{enumerate}
    \item \textbf{CNOT + Transpiler:} Circuit using CNOT basis gates, 
transpiled with optimization level 3.
    \item \textbf{ECR + Transpiler:} Circuit using ECR basis gates, 
transpiled with optimization level 3.
    \item \textbf{ECR (Topology-Aware) + Transpiler:} Circuit initially 
designed considering device connectivity using ECR gates, then transpiled 
with optimization level 3.
    \item \textbf{ECR (Topology-Aware) + Transpiler + Fixed Mapping:} 
Topology-aware ECR circuit, transpiled with optimization level 3 and a 
fixed 
logical-to-physical qubit mapping.
    \item \textbf{ECR (Topology-Aware) + Fixed Mapping (No Opt.):} 
Topology-aware ECR circuit with fixed mapping, bypassing subsequent 
transpiler optimization.
\end{enumerate}

As detailed in Table \ref{table:transpilation_11_qubits}, the results for 
the $11$-qubit system demonstrate a general performance advantage for 
ECR-based implementations over CNOT-based ones for both measurement 
protocols. For the short measurement scheme, the standard ECR 
implementation 
processed by the transpiler (Method 2) yields the highest accuracy 
($55.0\%$). However, a more pronounced effect is observed for the XOR 
measurement protocol. Employing a topology-aware circuit design combined 
with fixed qubit mapping (Methods 4 and 5) results in a significant 
accuracy 
improvement, achieving approximately $71.5\% - 71.8\%$, nearly a $20\%$ 
absolute increase compared to the standard CNOT transpiled approach 
($48.5\%$).

\begin{table}[h!]
\centering
\begin{tabular}{|l|c|c|}
\hline
\diagbox{Transpilation Strategy}{Measurement} & Short & XOR \\ \hline
CNOT + Transpiler & $43.3\%$ & $48.5\%$ \\ \hline
ECR + Transpiler & $55.0\%$ & $54.5\%$ \\ \hline
ECR (Topol.) + Transpiler & $36.1\%$ & $47.2\%$ \\ \hline
ECR (Topol.) + Transpiler + Fixed Map & $32.0\%$ & $71.5\%$ \\ \hline
ECR (Topol.) + Fixed Map (No Opt.) & $33.4\%$ & $71.8\%$ \\ \hline
\end{tabular}
\caption{Accuracy of different transpilation strategies for discrimination 
scheme on $11$-qubit obtained from on IBM Brisbane  using short and XOR 
measurement schemes. Each circuit was executed with 10,000 shots. Ambiguous 
measurement outcomes were randomly assigned.}
\label{table:transpilation_11_qubits}
\end{table}

It is worth observing that the substantial accuracy gain observed for the 
$11$-qubit XOR measurement using hardware-aware design and fixed mapping 
highlights the potential benefits of tailoring circuits to specific device 
characteristics. However, this approach presents a significant practical 
challenge: it currently necessitates manual, device-specific, and 
qubit-count-specific circuit construction and mapping. This process lacks 
straightforward algorithmic automation and requires considerable expert 
intervention for each target configuration, limiting its scalability and 
general applicability.

\subsection{Results}
\begin{figure}[!htp]
    \begin{minipage}{0.45\textwidth}
        \centering
        
\includegraphics[width=\linewidth]{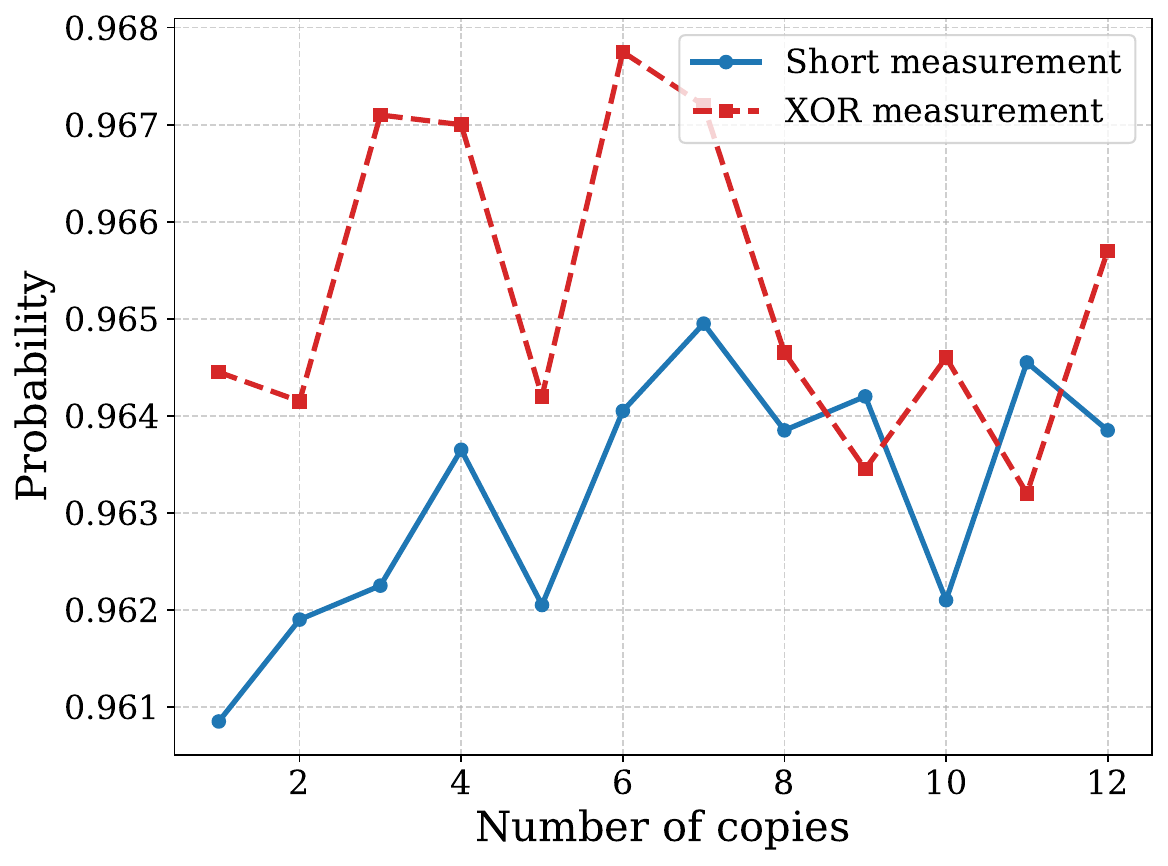}
        \caption*{(a) Purely sequential scheme of discrimination.}
        \label{plot:sequantial_unit}
    \end{minipage}
    \hfill
    \begin{minipage}{0.45\textwidth}
        \centering
        
\includegraphics[width=\linewidth]{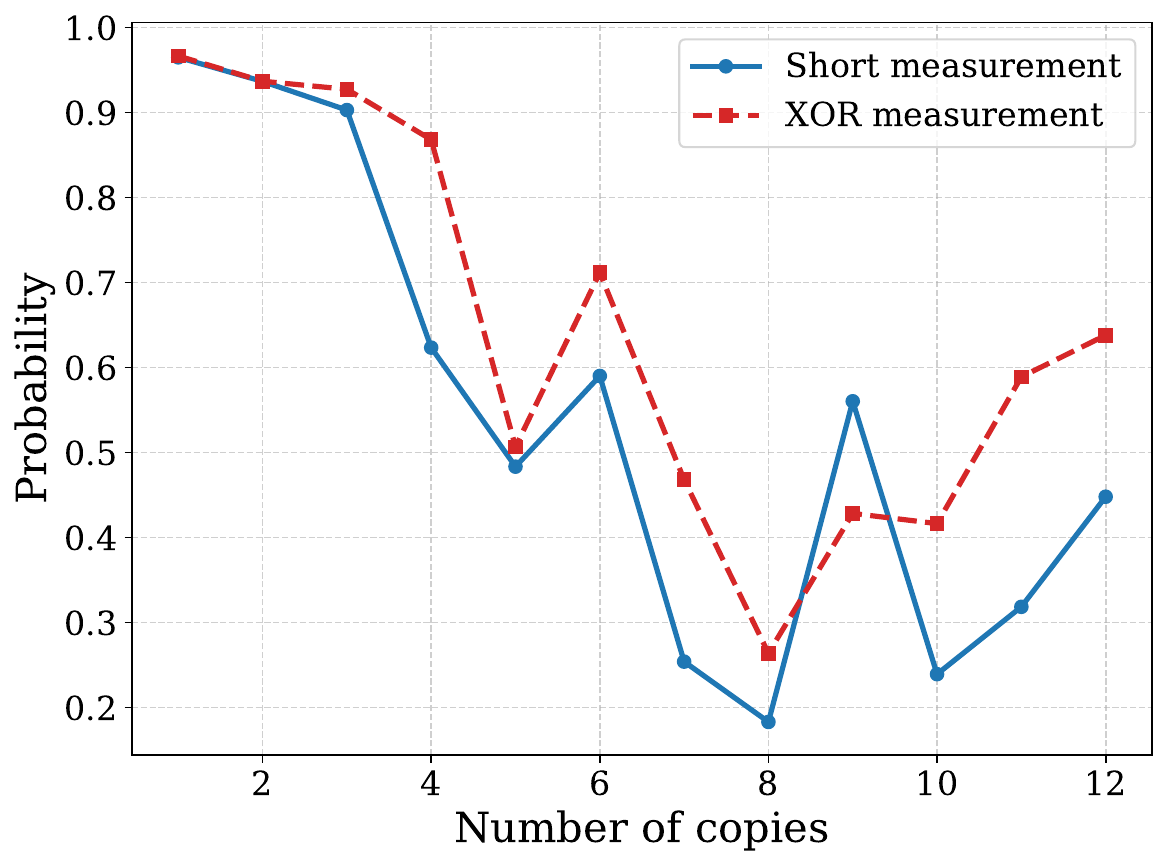}
        \caption*{(b) Purely parallel scheme of discrimination. }
        \label{plot:parallel_unit}
    \end{minipage}
    \caption{Probability of successful discrimination between the identity 
operation and the $\text{RZ}(\pi/ N)$ gate, where $N$ is number of copies 
of 
unknown unitary. The dashed red line corresponds to the XOR-based 
measurement strategy, while the solid blue line represents the short 
measurement scheme. Figure (a) illustrates the performance of the purely 
sequential scheme, and Figure (b) corresponds to the purely parallel 
scheme. 
Each circuit was executed with 10,000 shots per gate. Measurement outcomes 
that could not be unambiguously associated with either gate were randomly 
assigned to one of the two possible answers.}
    \label{fig:results_pure_unit}
\end{figure}

\begin{figure}[!htp]
    \begin{minipage}{0.45\textwidth}
        \centering
        
\includegraphics[width=\linewidth]{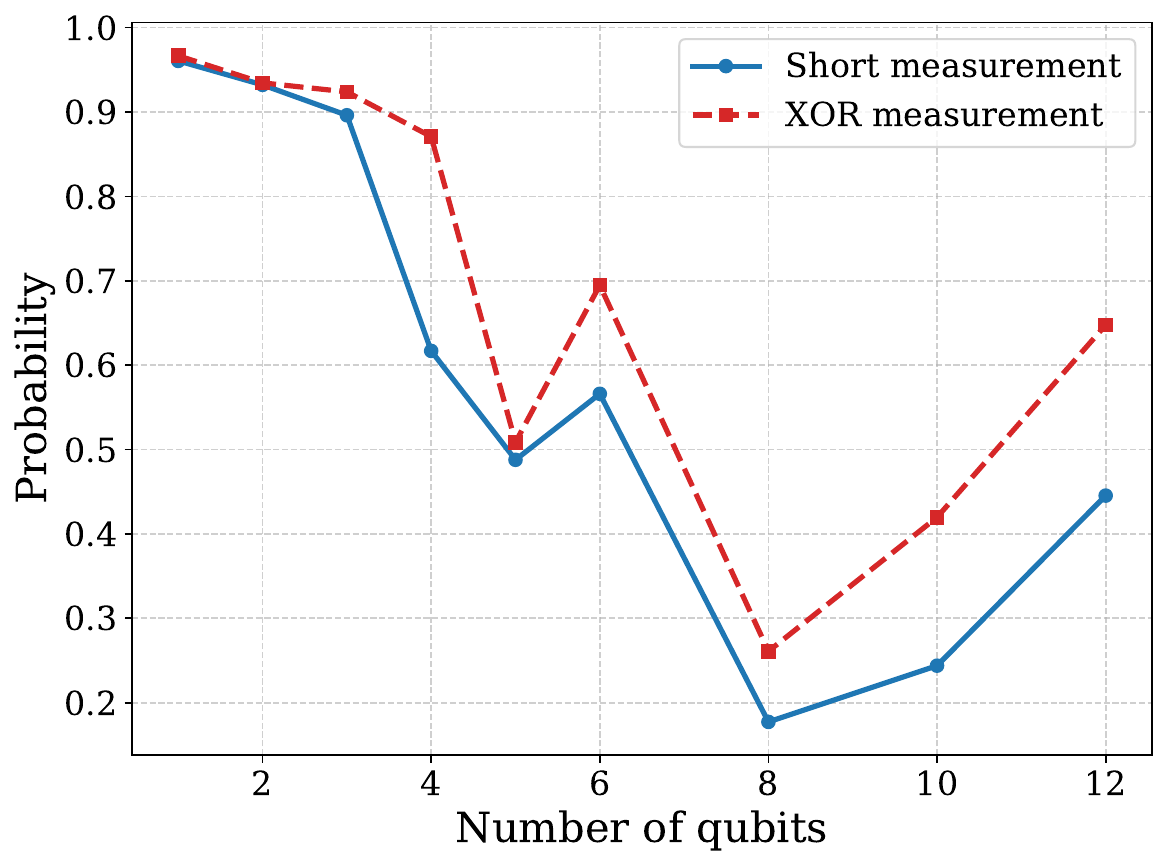}
        \caption*{(a) Hybrid scheme for $N =120$.}
    \end{minipage}
    \hfill
    \begin{minipage}{0.45\textwidth}
        \centering
        
\includegraphics[width=\linewidth]{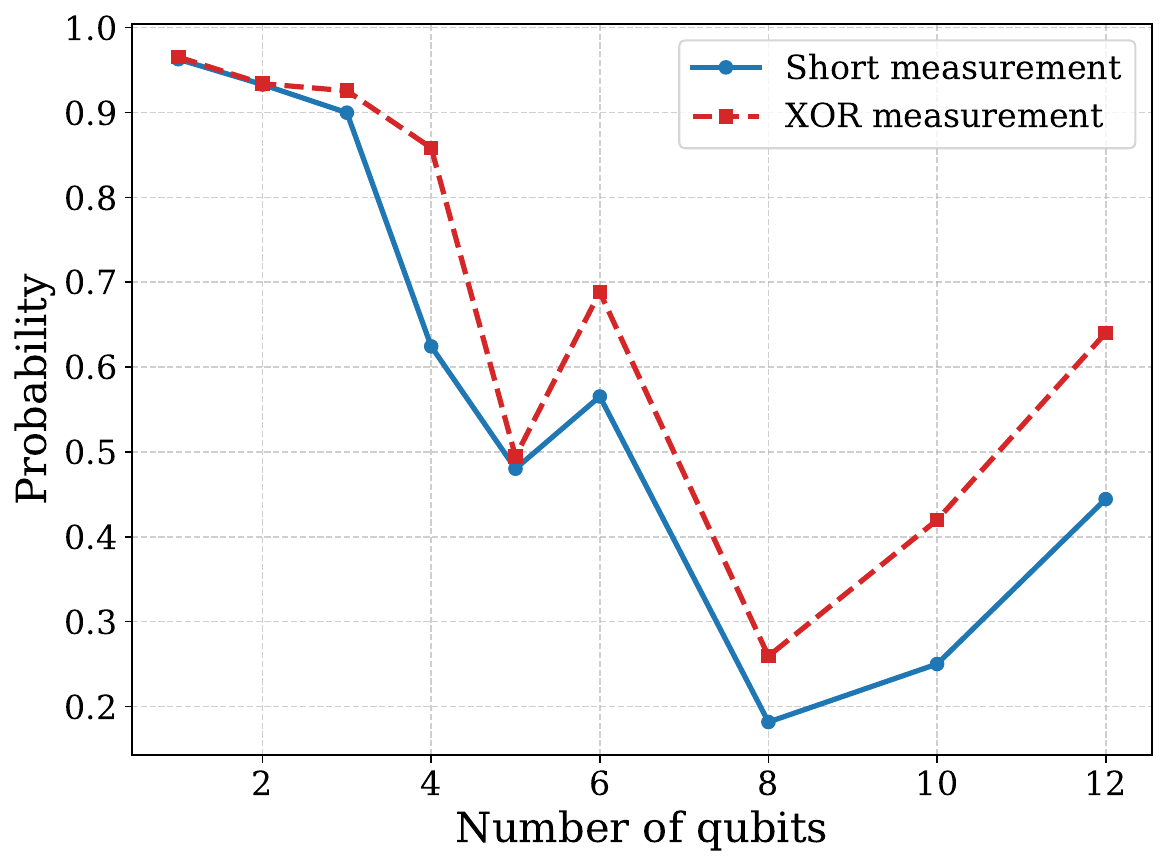}
        \caption*{(b) Hybrid scheme for $N=240$.}
    \end{minipage}
    \vspace{1em}
    \begin{minipage}{0.45\textwidth}
        \centering
        
\includegraphics[width=\linewidth]{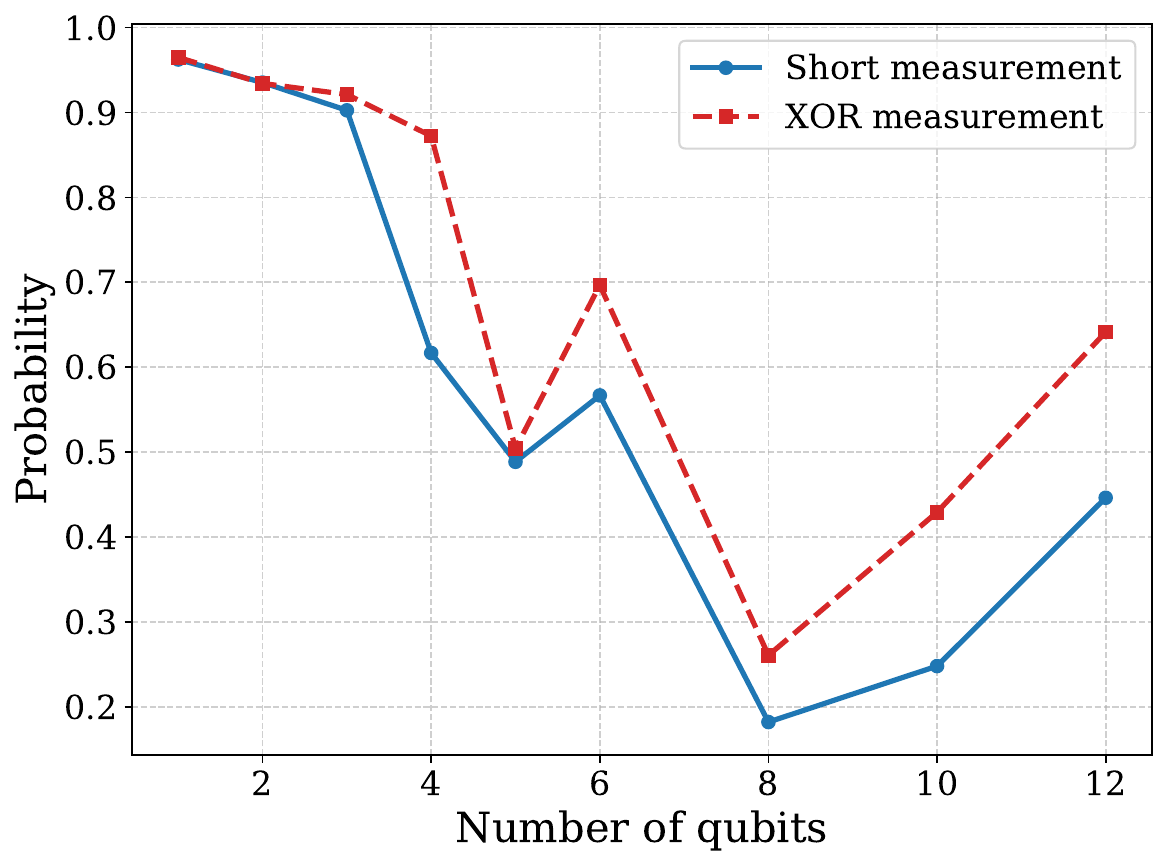}
        \caption*{(c) Hybrid scheme for $N=1200$.}
    \end{minipage}
    \caption{Probability of successful discrimination between the identity 
operation and the \text{RZ}($\pi/N$) gate, where $N$ is number of
copies of unknown unitary as a function of the width of hybrid rectangular 
scheme. The dashed red line corresponds to the XOR-based measurement 
strategy, while the solid blue line represents the short measurement 
scheme. 
Figure (a) illustrates the performance of hybrid scheme with 120 copies in 
total, Figure (b) corresponds to hybrid scheme with 240 copies in total, 
and 
Figure (c) corresponds to hybrid scheme with 1200 copies in total. Each 
circuit was executed with 10,000 shots per gate. Measurement outcomes that 
could not be unambiguously associated with either gate were randomly 
assigned to one of the two possible answers.}
\label{fig:results_hybrid_unit}
\end{figure}

From the initial experiments involving purely sequential and purely 
parallel discrimination schemes (see Fig.~\ref{fig:results_pure_unit}), we 
observe that the use of entangling gates on real quantum devices introduces 
a greater error overhead than the decoherence effects arising from 
increased 
gate depth in sequential protocols.
To further investigate this observation, we conducted a series of tests 
using hybrid rectangular (sequentially-parallel) schemes with a fixed 
number 
of unknown gate applications (see Fig.~\ref{fig:results_hybrid_unit}). The 
results consistently show a significant increase in error rates as more 
entangling gates are introduced. This trend reinforces the conclusion that 
the primary source of performance degradation in current devices stems from 
the imperfections of multi-qubit gate operations rather than decoherence 
from circuit depth alone. Another notable observation is that different 
measurement schemes have a visible but not dominant impact on the overall 
error rate, further suggesting that the circuit width and the number of 
entangled qubits are the primary contributors to performance degradation. 
This reinforces the conclusion that the main source of error arises from 
the 
entanglement rather than from the specifics of the measurement strategy.

\subsection{Discussion}
During the experiments, we identified a specific issue with the IBM Q 
platform namely measurements in experiments involving entanglement of 
specific number, typically five or more, qubits exhibited random bit-flip 
errors. This type of error is evident in Fig.~\ref{fig:results_pure_unit} 
and Fig.~\ref{fig:results_hybrid_unit} and appears to affect the 
measurement 
results of all qubits at the same time independent of the measurement 
strategy. In particular, these anomalies can be removed simply by swapping 
the sets of expected answers, effectively compensating for the observed 
flips and restoring consistency in the results. 
To ensure that the issue was not caused by errors in our own 
implementation, we performed several verification steps. First, the 
experiments were repeated multiple times on different dates. Second, the 
final transpiled circuits were simulated using a noiseless simulator to 
confirm the expected theoretical behavior. Third, we applied M3 error 
mitigation techniques, which had no effect on overall results. Given that 
the observed errors persist across all experiments, regardless of the RZ 
gate angle or the measurement strategy employed, we hypothesize the 
presence 
of a systematic hardware or software artifact specific to the IBMQ 
platform, 
particularly the Brisbane quantum device.  Further investigation and 
hypothesis-driven testing are required to understand this behavior.

\begin{figure}[!htp]
    \begin{minipage}{0.45\textwidth}
        \centering
        
\includegraphics[width=\linewidth]{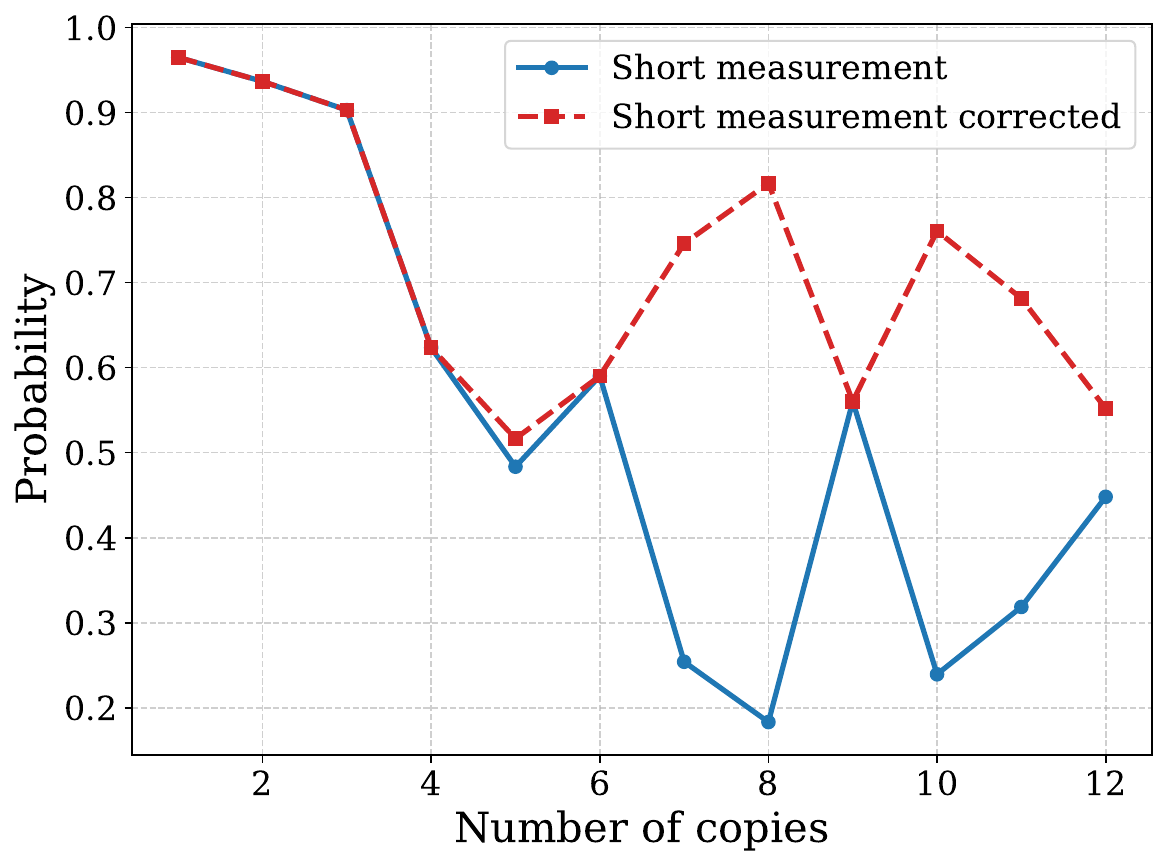}
        \caption*{(a) Usage of the short measurement.}
        \label{plot:parallel_short_corrected}
    \end{minipage}
    \hfill
    \begin{minipage}{0.45\textwidth}
        \centering
        
\includegraphics[width=\linewidth]{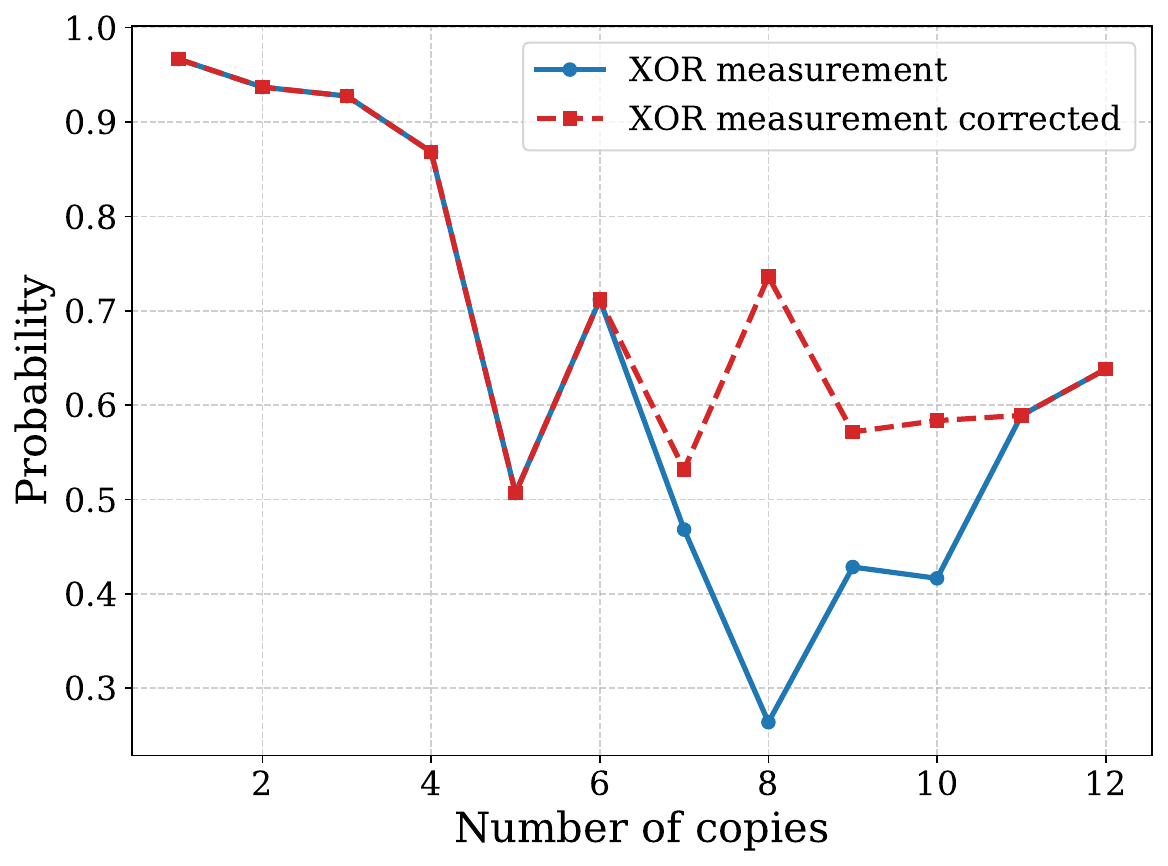}
        \caption*{(b) Usage of the XOR measurement.}
        \label{plot:parallel_xor_corrected}
    \end{minipage}
    \caption{Probability of successful discrimination between the identity 
gate and the \text{RZ}$(\pi/N)$  gate, where $N$ is number of copies of 
unknown unitary  
    as a function of the number of channel uses in purely parallel scheme, 
after applying post-processing correction. The dashed red line corresponds 
to the corrected measurement strategy, where the sets of answers associated 
with each channel were swapped in cases where the success probability 
dropped below 0.5. The solid blue line shows the original results using the 
short measurement strategy. Figure (a) presents the performance of the 
short 
measurement scheme, while Figure (b) displays the XOR-based scheme. Each 
circuit was executed with 10,000 shots per gate. Measurement outcomes that 
could not be unambiguously attributed to either gate were randomly assigned 
to one of the two possible answers.}
    \label{fig:results_corrected}
\end{figure}

\newpage

\section{Discrimination of unitary channels on IBM-Q with 
processing}\label{sec:reali2}

In this section, we will present the second example of discrimination task 
with a different pair of unitary channels and run it on IBM Quantum 
processor Brisbane. 
\subsection{Example 2}
 In the second experiment, we take the following unitary channels to 
discriminate $\Phi_U$ for $U = 
\sqrt{X}\mathrm{RZ}(\frac{-\pi}{2N})\sqrt{X}$ 
and $\Phi_V$ for $V = \sqrt{X}\mathrm{RZ}(\frac{\pi}{2N})\sqrt{X}$, where 
$\sqrt{X} = \frac{1}{2} \begin{pmatrix}
     1+i & 1-i\\
     1-i & 1+i
 \end{pmatrix}$. It is easy to check that condition $\theta(V^\dagger U) =  
\frac{\pi}{N}$ is satisfied. It implies that for $N$-shot discrimination 
scenario, we will achieve perfect discrimination. Let us fix $N = wd$, 
where 
$w,d \in \mathbb{N}$. In this setup, instead of mid-processing $X_i = 
(V^\dagger)^{\otimes w}$ we use hardware-friendly processing $X_i = 
X^{\otimes w}$. For convince we add also pre-processing unitary operation 
$X_0 = (X\sqrt{X})^{\otimes w}$ and post-processing operation $X_d = 
(\sqrt{X}X)^{\otimes w}$. Combining processing operators with the black-box 
$\Phi_U$ we get the following unitary operation
 \begin{equation}
     U_* = X_dU^{\otimes w}X_{d-1}\cdots X_1U^{\otimes w}X_0 = 
\left(\mathrm{RZ}(\frac{-\pi}{2N})^d\right)^{\otimes w}.
 \end{equation}
 Similarly, for $\Phi_V$ the combined unitary circuit equals $V_* = 
\left(\mathrm{RZ}(\frac{\pi}{2N})^d\right)^{\otimes w}$.
 Note that $\theta(\left(\mathrm{RZ}(\frac{-\pi}{2N})^d\right)^{\otimes 
w}\left(\mathrm{RZ}(\frac{-\pi}{2N})^d\right)^{\otimes w}) = 
wd\theta(\mathrm{RZ}(\frac{-\pi}{N})) = N\frac{\pi}{N} = \pi$, so in 
theory, 
each shape $w,d$ gives perfect discrimination. As $X, \sqrt{X}, 
\mathrm{RZ}$ 
are native gates in IBMQ Brisbane this part of the circuit is implemented 
exactly as stated. To define the discriminator $\ket{\psi}$, we solve
 \begin{equation}
     \bra{\psi} \left(\mathrm{RZ}(\frac{-\pi}{N})^d\right)^{\otimes 
w}\ket{\psi} = 0.
 \end{equation}
We find that the input state can be taken as the GHZ state $\ket{\psi} = 
\frac{1}{\sqrt{2}}(\ket{0\ldots0} + \ket{1\ldots1})$, which can be 
implemented by Hadamard gate followed by cascade of control-X gates. We let 
the IBM transpiler to optimize the input state circuit with the 
optimization 
level set to $3$.

We distinguish two situations depending on the value of $\Phi$. If $\Phi = 
\Phi_U$, then the input state $\ket{\psi}$ evolved under $U_*$ is equal to 
$\ket{\psi_{-i}} = \frac{1}{\sqrt{2}}(\ket{0\ldots0} - i \ket{1\ldots1})$. 
For $\Phi = \Phi_V$ we get after the evolution $V_*$ that $\ket{\psi_{i}} = 
\frac{1}{\sqrt{2}}(\ket{0\ldots0} + i \ket{1\ldots1})$. At the measurement 
stage we implement $\mathrm{RZ}(\pi/2)$ on the first qubit followed by 
parallel application of Hadamard gates $H^{\otimes w}$.
Each qubit is then measured in the $Z$-basis. We let the IBM transpiler to 
optimize the measurement circuit with the optimization level set to $3$.
At the measurement stage, after applying $\mathrm{RZ}(\pi/2)$ and 
$H^{\otimes w}$ we get
\begin{equation}
    H^{\otimes w}(\mathrm{RZ}(\pi/2) \otimes \Id)\ket{\psi_{\mp i}} = 
\ket{+\ldots+} \pm \ket{-\ldots-},
\end{equation}
where we used the notation of plus/minus states $\ket{+} = H\ket{0}, 
\ket{-} = H\ket{1}$. Observe that if $(b_1,\ldots,b_w)$ is a bit string we 
received from measuring $\ket{+\ldots+} \pm \ket{-\ldots-}$, then in 
theory, 
$\Phi = \Phi_U$ if and only if $b_1 \oplus \cdots \oplus b_w \equiv 0$. 
Similarly, $\Phi = \Phi_V$ if and only if $b_1 \oplus \cdots \oplus b_w 
\equiv 1$.

 \subsection{Results}
From relatively small number of copies ($N=4,16,32$) of unitary channels to 
be discriminated, the experiments involving purely sequential schemes are 
preferable (see Fig.~\ref{fig:results_processing_sequential}).  We observe 
that the use of entangling gates on real quantum devices introduces a 
higher 
error overhead than the decoherence effects arising from increased gate 
depth in sequential protocols.  For an increasing number of copies ($N=64, 
96, 1024$) of unitary channels, we can observe the advantage of usage 
sequentially-paralleled schemes to achieve more precise results (see 
Fig.~\ref{fig:results_processing_rectangular}). The results consistently 
show that after some threshold of gate composition, the existing 
decoherence 
or accumulative calibration imperfections gives higher error rates than the 
error rates used to create an entangled state.

\subsection{Discussion}
\begin{figure}[!htp]
    \begin{minipage}{0.45\textwidth}
        \centering
        
\includegraphics[width=1.1\linewidth]{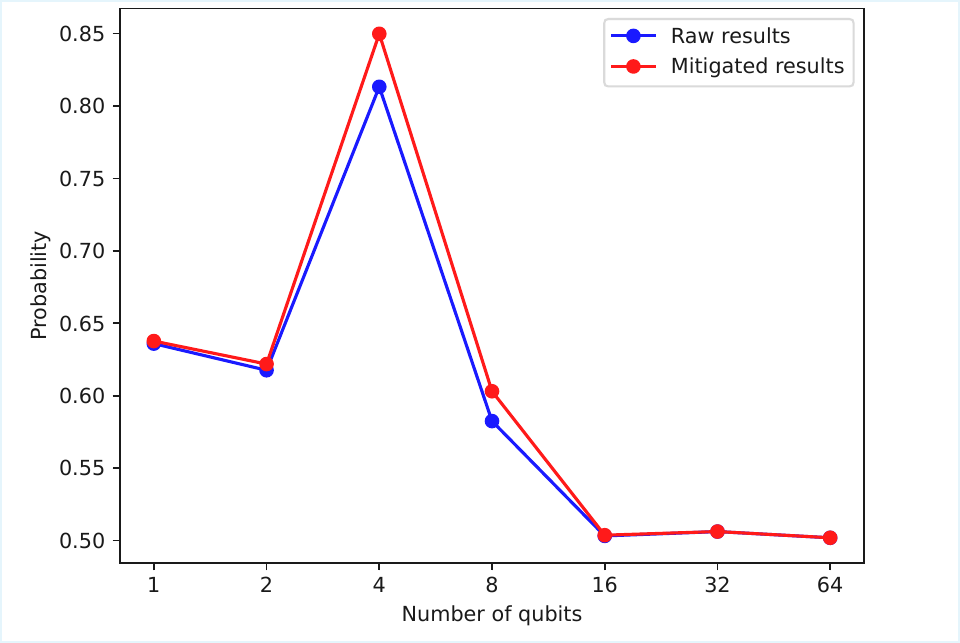}
        \caption*{(a) Hybrid scheme for $N =64$.}
    \end{minipage}
    \hfill
    \begin{minipage}{0.45\textwidth}
        \centering
        
\includegraphics[width=1.1\linewidth]{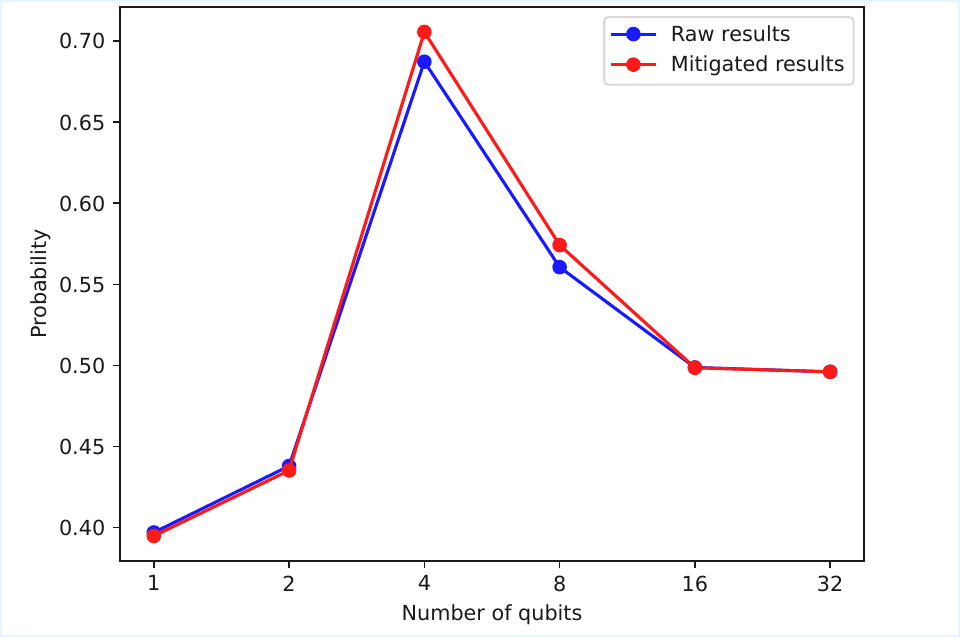}
        \caption*{(b) Hybrid scheme for $N=96$.}
    \end{minipage}
    \vspace{1em}
    \begin{minipage}{0.45\textwidth}
        \centering
        
\includegraphics[width=1.1\linewidth]{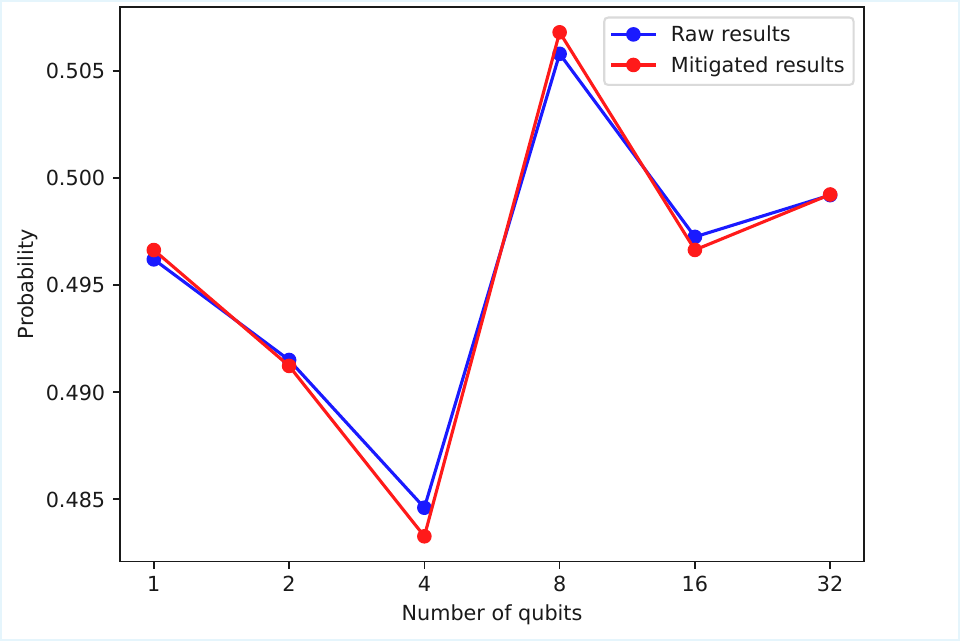}
        \caption*{(c) Hybrid scheme for $N=1024$.}
    \end{minipage}
    \caption{Probability of successful discrimination between the unitary 
operator $U = \sqrt{X} \mathrm{RZ}(\frac{-\pi}{2N}) \sqrt{X}$ and $V = 
\sqrt{X} \mathrm{RZ}(\frac{\pi}{2N}) \sqrt{X}$, where $N$ is number of
copies of unknown unitary as a function of the width of hybrid rectangular 
scheme using the short measurement. The blue line corresponds to the 
results 
obtained directly from IBM Q processor Brisbane, while the red line 
represents the results after error mitigation using MThree package 
\cite{mthree, mthreepublication}. Figure (a) illustrates the performance of 
hybrid schemes with 64 copies in total, Figure (b) corresponds to hybrid 
schemes with 96 copies in total, and Figure (c) corresponds to hybrid 
schemes with 1024 copies in total. Each circuit was executed with 10,000 
shots per gate.}
    \label{fig:results_processing_rectangular}
\end{figure}

In the experiment with $N=1024$ copies of black-box we obtained totally 
random results for all schemes considered (see 
Fig.~\ref{fig:results_processing_rectangular})(c)). Sequential scheme 
required too much gate composition while parallel one required creating too 
big GHZ state. Both circuits introduced too much errors to obtain any 
notable results. The final question that we can ask is if we can do better 
than that by exploring \textit{suboptimal} circuits. A simple idea goes as 
follow. We make independent sequential experiments on each of $w$ qubits. 
For each, we compose $d$ quantum circuits and do independent measurements 
that indicate which black-box is preferable. As $\theta(V^\dagger U) = 
\frac{\pi}{N}$ for each qubit we get the maximum angle spread of $\theta = 
\frac{\pi}{N}d = \frac{\pi}{w} < \pi$. Hence, the protocol is suboptimal. 
The probability of successful discrimination can be boosted by using 
majority voting on the results collected from the quantum computer. Lets 
assume we collected $k$ times label $0$ indicating $\Phi_U$ and $w-k$ times 
label $1$ standing for $\Phi_V$. If $k > w-k$ we guess that $\Phi = \Phi_U$ 
(and for $k < w-k$ we guess $\Phi = \Phi_U$). In the case $k=w-k$ we take 
random guess. 

\begin{figure}[!htp]
    \begin{minipage}{0.45\textwidth}
        \centering
        
\includegraphics[width=1.1\linewidth]{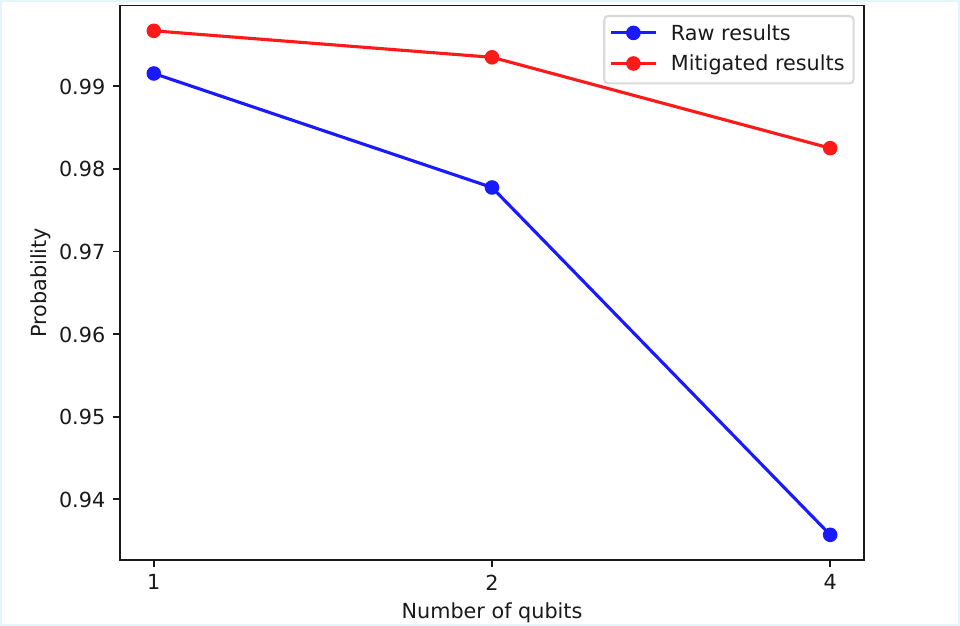}
        \caption*{(a) Hybrid scheme for $N =4$.}
    \end{minipage}
    \hfill
    \begin{minipage}{0.45\textwidth}
        \centering
        
\includegraphics[width=1.1\linewidth]{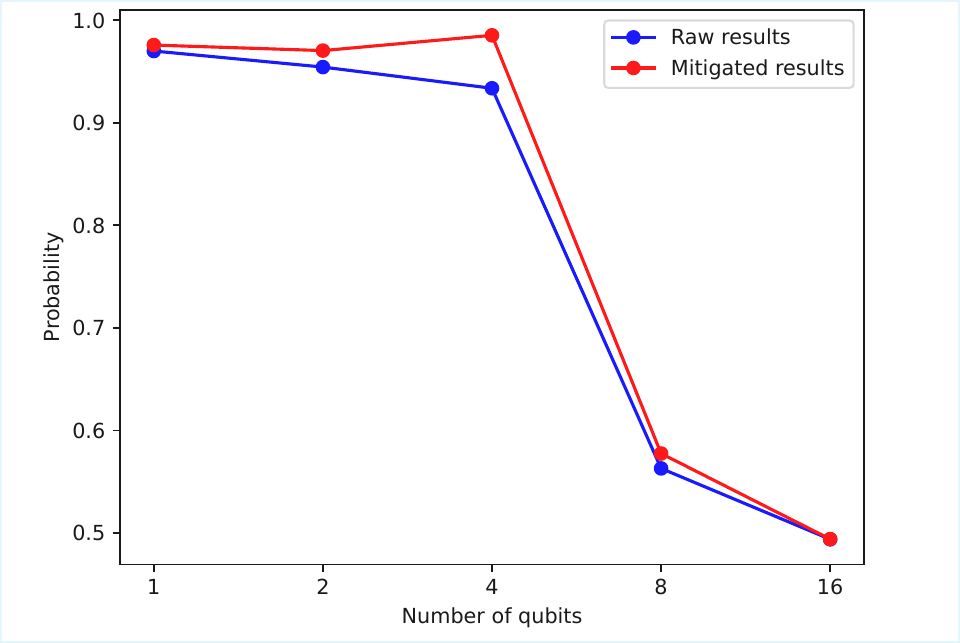}
        \caption*{(b) Hybrid scheme for $N=16$.}
    \end{minipage}
    \vspace{1em}
    \begin{minipage}{0.45\textwidth}
        \centering
        
\includegraphics[width=1.1\linewidth]{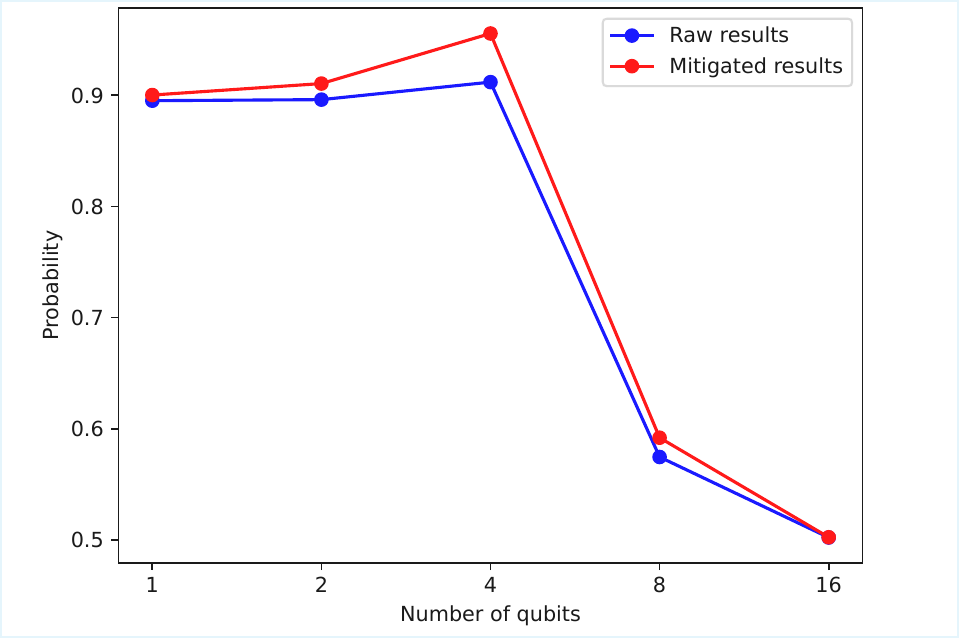}
        \caption*{(c) Hybrid scheme for $N=32$.}
    \end{minipage}
    \caption{Probability of successful discrimination between the unitary 
operator $U = \sqrt{X} \mathrm{RZ}(\frac{-\pi}{2N}) \sqrt{X}$ and $V = 
\sqrt{X} \mathrm{RZ}(\frac{\pi}{2N}) \sqrt{X}$, where $N$ is number of
copies of unknown unitary as a function of the width of hybrid rectangular 
scheme using the short measurement. The blue line corresponds to the 
results 
obtained directly from IBM Q processor Brisbane, while the red line 
represents the results after error mitigation using MThree package 
\cite{mthree, mthreepublication}. Figure (a) illustrates the performance of 
hybrid schemes with 4 copies in total, Figure (b) corresponds to hybrid 
schemes with 16 copies in total, and Figure (c) corresponds to hybrid 
schemes with 32 copies in total. Each circuit was executed with 10,000 
shots 
per gate.}
    \label{fig:results_processing_sequential}
\end{figure}

We applied the following suboptimal procedure for $N=1024$. We used $w=32$ 
qubits in one experiment and on each of them, we composed $d=32$ times the 
operation $\Phi$. According to Fig.~\ref{fig:results_processing_sequential} 
we should expect around $90\%$ accuracy for each qubit. In combine we 
received $p_{\mathrm{succ}} = 0.56765$ that is better than any optimal 
scheme. Similarly, we did the experiment for $N = 96$ and $w=3, d=32$ 
resulting in $p_{\mathrm{succ}} = 0.74685$, which is better result than 
indicated by Fig.~\ref{fig:results_processing_rectangular})(b). 

Can we say that suboptimal circuits are in favor for unitary channel 
discrimination? The answer is not straightforward. For example, if $N = 
64$ 
then the best result from Fig.~\ref{fig:results_processing_rectangular})(a) 
is slightly better than suboptimal procedure presented above even in the 
absence of the noise, $p_{\mathrm{succ}} = 0.85295$.

\section{Conclusion}\label{sec:conclusion}
In this work, we have studied the discrimination of two quantum unitary 
channels and benchmarked various schemes for perfect discrimination between 
them. The benchmarks have been performed using the IBM Brisbane quantum 
device. As the first example,  we have chosen the discrimination task 
between the identity operation and the \text{RZ}$(\phi)$ gate with no 
processing between them. 
To optimize circuit performance for this discrimination task, we also 
evaluated different quantum circuit transpilation approaches on 6- and 
11-qubit subsystems, comparing the circuit implementation with CNOT and ECR 
gates and the impact of manual qubit mapping. Although fixed qubit mapping 
did not significantly improve accuracy in the smaller 6-qubit system (see 
Table~\ref{table:transpilation_6_qubits}), topology-aware circuit design 
combined with fixed mapping yielded substantial gains on the 11-qubit 
system 
when using XOR-based measurements, underscoring the importance of 
hardware-aware optimization for larger circuits (see 
Table~\ref{table:transpilation_11_qubits}).
In this experiment, we find the trend that deeper circuit architectures, 
which minimize entanglement overhead while preserving discrimination power, 
are significantly more resilient to hardware noise (see 
Fig.~\ref{fig:results_pure_unit}). Our preliminary research suggested the 
claim that algorithm designers should prioritize depth over width in 
circuit 
construction if possible. In the second example, 
we have considered the discrimination between unitaries $U = 
\sqrt{X}\mathrm{RZ}(\frac{-\pi}{2N})\sqrt{X}$ and $\Phi_V$ for $V = 
\sqrt{X}\mathrm{RZ}(\frac{\pi}{2N})\sqrt{X}$
with processing being composition of X and $\sqrt{X}$ gates. There, we have 
observed 
the advantage of
usage sequentially-paralleled schemes to achieve more precise results.
The results consistently show that after some threshold of gate 
composition, the existing decoherence or accumulative calibration 
imperfections gives higher error
rates than the error rates used to create an entangled state.

Circuit geometries beyond square layouts may offer a more accurate 
reflection of the capabilities of the device. This result can be applied to 
various black-box tasks with many copies, such as quantum phase 
estimation~\cite{dorner2009optimal}. In that case, discrimination schemes 
that are theoretically suboptimal achieve good experimental 
performance~\cite{svore2013faster}.

\section*{Data Availability Statement}
The data that support the findings of this study are openly available in 
\href{https://github.com/Dotnester/experimental_study_of_multiple_shot_channel_discrimination}{Github
 repository} at 
\url{https://github.com/Dotnester/experimental_study_of_multiple_shot_channel_discrimination}
 and on Zenodo~\cite{bilek_2025_15464711}.

\section*{Acknowledgements}

It is a pleasure to thank Łukasz Pawela for numerous discussions concerning 
experiments on IBM quantum devices.

AB is supported by Grant of SGS No. SP2025/049, VŠB - Technical University 
of Ostrava, Czech Republic.
JH is supported by the Ministry of Education, Youth and Sports of the Czech 
Republic through the e-INFRA CZ (ID:90254), with the financial support by 
Grant of SGS No. SP2025/049, VŠB - Technical University of Ostrava, Czech 
Republic. 
TB is supported by Grant of SGS No. SP2025/049, VŠB - Technical University 
of Ostrava, Czech Republic. 
  PL is supported by the Ministry of Education, Youth and Sports of the 
Czech Republic through the e-INFRA CZ (ID:90254), with the financial 
support 
of the European Union under the REFRESH – Research Excellence For 
REgionSustainability and
High-tech Industries project number CZ.10.03.01/00/22\_003/0000048 via the 
Operational Programme Just Transition.  
RK is supported by the European Union under the REFRESH – Research 
Excellence For REgionSustainability and
High-tech Industries project number CZ.10.03.01/00/22\_003/0000048 via the 
Operational Programme Just Transition.

\bibliographystyle{unsrt} 
\bibliography{seq_channel_discrim.bib}
\end{document}